\documentclass[prl,showpacs]{revtex4}
\usepackage{graphicx,psfrag,amssymb,amsmath}

\begin{document}
\title{Unconventional charge density wave in the organic conductor $\alpha$-(BEDT-TTF)$_2$KHg(SCN)$_4$}
\author{Kazumi Maki}
\affiliation{Department of Physics and Astronomy, University of Southern
California, Los Angeles CA 90089-0484, USA} 
\author{Bal\'azs D\'ora}
\affiliation{The Abdus Salam ICTP, Strada Costiera 11, I-34014, Trieste, Italy} 
\author{Mark Kartsovnik}
\affiliation{Walther-Meissner Institute, D-85748 Garching, Germany}
\author{Attila Virosztek}
\affiliation{Department of Physics, Budapest University of Technology and 
Economics, H-1521 Budapest, Hungary}
\affiliation{Research Institute for Solid State Physics and Optics, P.O.Box
49, H-1525 Budapest, Hungary}
\author{Bojana Korin-Hamzi\'c}
\affiliation{Institute of Physics, POB 304, HR-10001 Zagreb, Croatia}
\author{Mario Basleti\'c}
\affiliation{Department of Physics, Faculty of Science, POB 331, HR-10001
Zagreb, Croatia}

\date{\today}

\begin{abstract}
The low temperature phase (LTP) of $\alpha$-(BEDT-TTF)$_2$KHg(SCN)$_4$
salt is known for its surprising angular dependent magnetoresistance (ADMR), which has been
studied intensively in the last decade. However, the nature of the LTP has not been understood
until now. Here we analyse theoretically ADMR in unconventional (or nodal) charge density wave (UCDW).
In magnetic field the quasiparticle spectrum in UCDW is quantized, which gives rise
to spectacular ADMR. The present model accounts for many striking
features of ADMR data in $\alpha$-(BEDT-TTF)$_2$KHg(SCN)$_4$.
\end{abstract}


\maketitle

The series of quasi-two dimensional organic conductors $\alpha$-(BEDT-TTF)$_2$MHg(SCN)$_4$
(where BEDT-TTF denotes bis(ethylenedithio)tetrathiafulvalene and M=K, NH$_4$, Rb and Tl)
have attracted considerable attention over the last few years due to two different ground states and rich phenomena
associated with them\cite{singl}.

Whereas the M=NH$_4$ compound becomes superconducting below $1.5$K, other salts enter at $T_c=8-12$K into a specific
low temperature phase (LTP) with striking ADMR.
From the magnetic phase diagram of LTP it is now believed that LTP in not SDW but
a kind of CDW, though no detailed characterization is available\cite{andres}.
We have proposed recently that unconventional (or nodal) charge density wave (UCDW) can account for a number of features
in LTP in $\alpha$-(BEDT-TTF)$_2$KHg(SCN)$_4$ including the threshold electric field\cite{kuszobter,rapid,tesla,epladmr,nesting}.
Recently UCDW and USDW have been proposed by several authors as possible electronic ground state in quasi-one
dimensional and quasi-two dimensional crystals\cite{Ners1,Ners2,benfatto,nagycikk,nayak}. Unlike conventional DW\cite{gruner}, the
order parameter in UCDW $\Delta(\bf k)$
depends on the quasiparticle wave vector $\bf k$.
In $\alpha$-(BEDT-TTF)$_2$KHg(SCN)$_4$ salts, where the conducting plane lies in the a-c plane and the quasi-one
dimensional Fermi surface is perpendicular to the a-axis, we assume that $\Delta({\bf k})=\Delta\cos(ck_z)$ or 
$\Delta\sin(ck_z)$ (i.e.
$\Delta(\bf k)$ depends on $\bf k$ perpendicular to the most conducting direction), where $c=9.778$\AA $ $ is the lattice
constant along the c-axis\cite{mori}. 
It is known also that the thermodynamics of UCDW and USDW is practically the same as the one in d-wave
superconductor\cite{nagycikk,d-wave}.
Also in spite of the clear thermodynamic signal, the first order terms in $\Delta(\bf k)$ usually vanishes when averaged
over the Fermi surface. This implies neither clear x-ray signal for UCDW, nor spin signal for USDW.
Due to this fact unconventional density waves are sometimes called the phase with hidden order parameter\cite{nayak}.

\begin{figure}[h]
\psfrag{x}[t][b][1.2][0]{$B$(T)}
\psfrag{y}[b][t][1.2][0]{$R$(Ohm)}
\psfrag{m1}[t][b][1][0]{$T=1.4$K}
\psfrag{m2}[][][1][0]{$T=4.14$K}
\includegraphics[width=7cm,height=7cm]{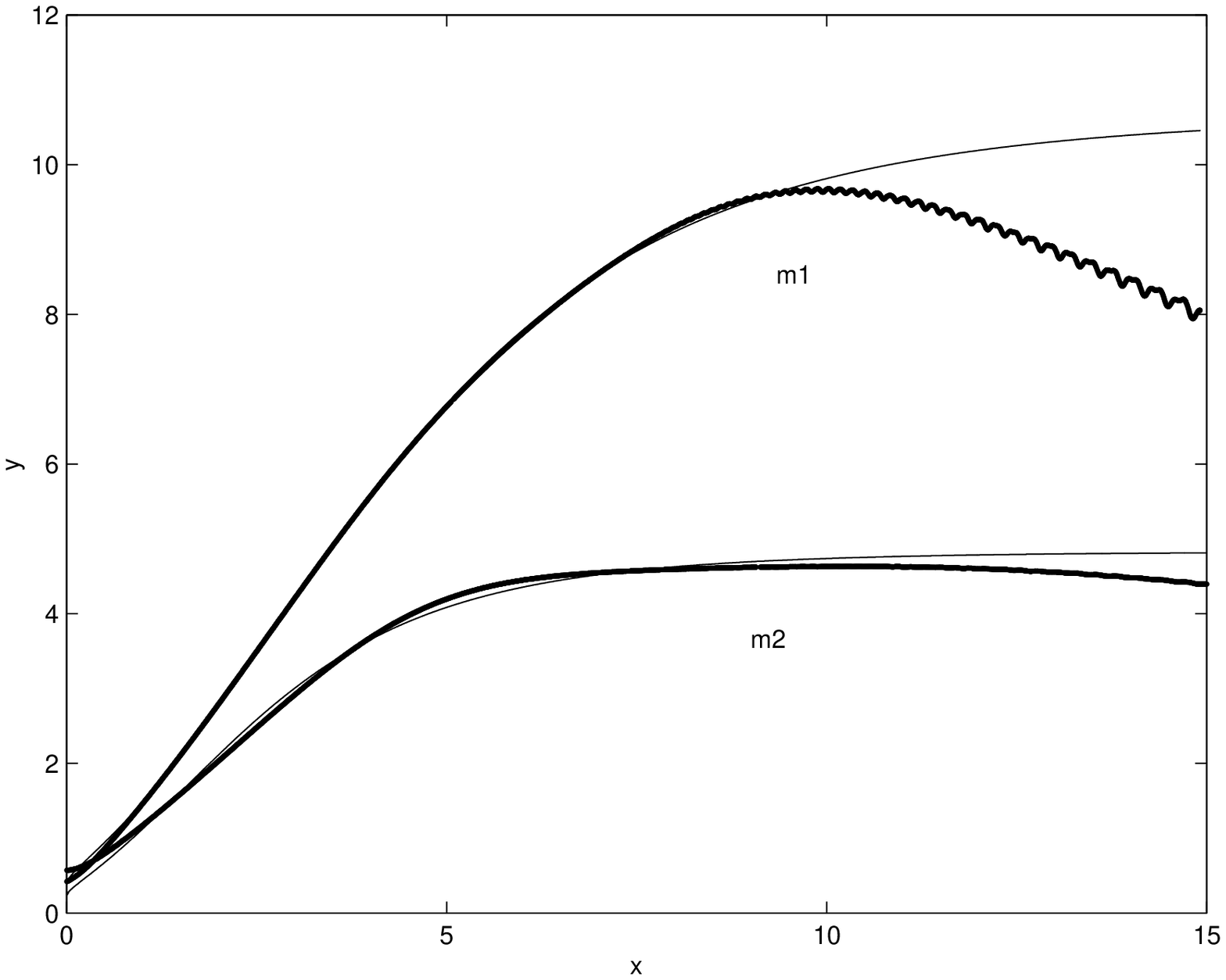}
\psfrag{B}[bl][tr][1][0]{$\bf B$}
\psfrag{c}[b][t][1][0]{$c$}
\psfrag{a}[b][t][1][0]{$a$}
\psfrag{b}[r][l][1][0]{$b$}
\psfrag{pp}[][][1][0]{$\phi$}
\psfrag{p}[][][1][0]{$\theta$}

\vspace*{-6.5cm}\hspace*{-3.7cm}
\includegraphics[width=2cm,height=2cm]{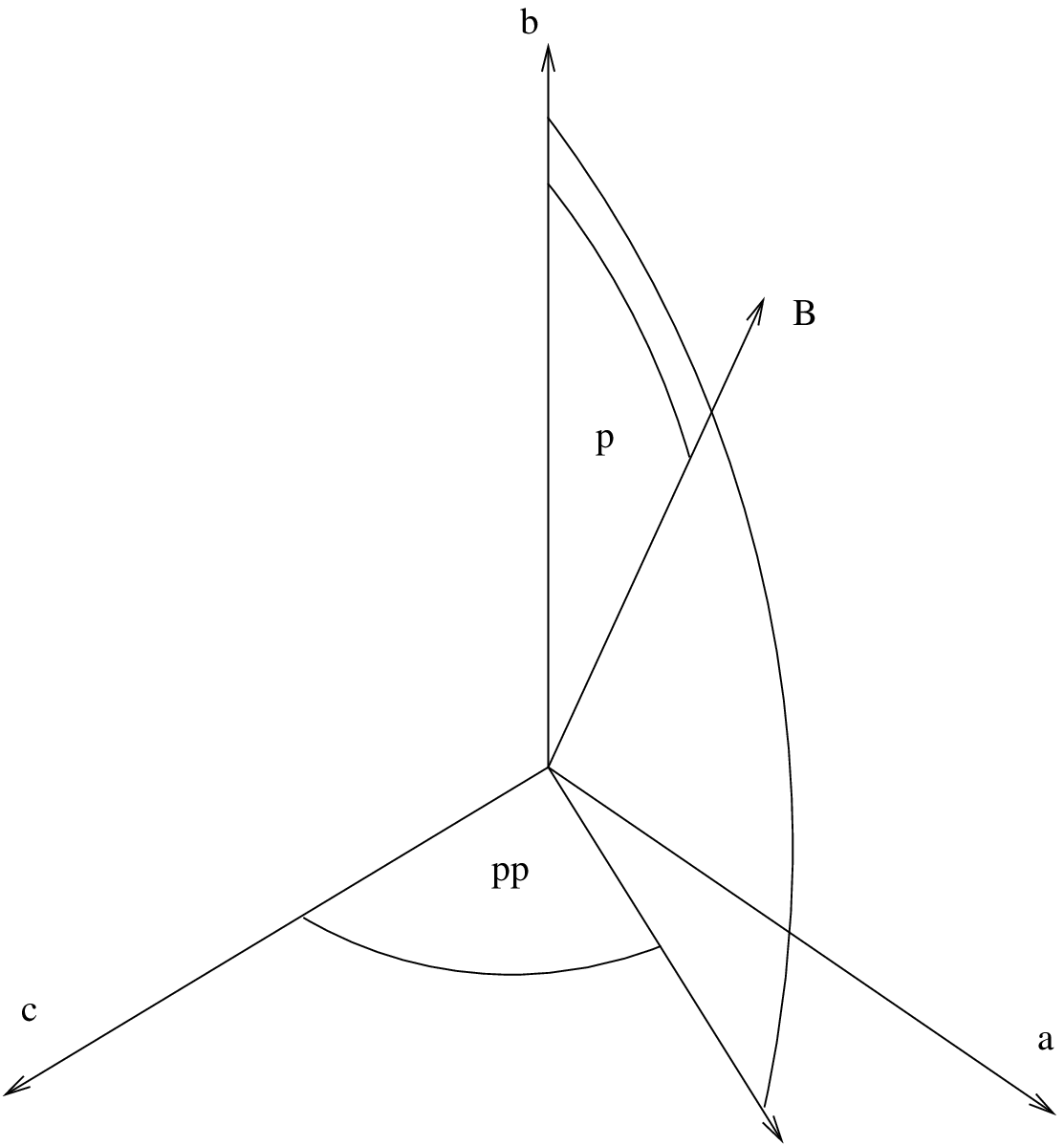}

\vspace*{5cm}
\caption{The magnetoresistance is plotted for $T=1.4$K and $4.14$K as a function of magnetic field. The thick solid is the
experimental data, the thin one denotes our fit based on Eq. (\ref{fit}). The inset shows the geometrical
configuration of the magnetic field with respect to
 the conducting plane.}\label{fig:koord}
\end{figure}

\begin{figure}[h]
\psfrag{x}[t][b][1.2][0]{$T$(K)}
\psfrag{y}[b][t][1.2][0]{$R$(Ohm)} 
\psfrag{m3}[l][r][1][0]{$B=15$T}
\includegraphics[width=7cm,height=7cm]{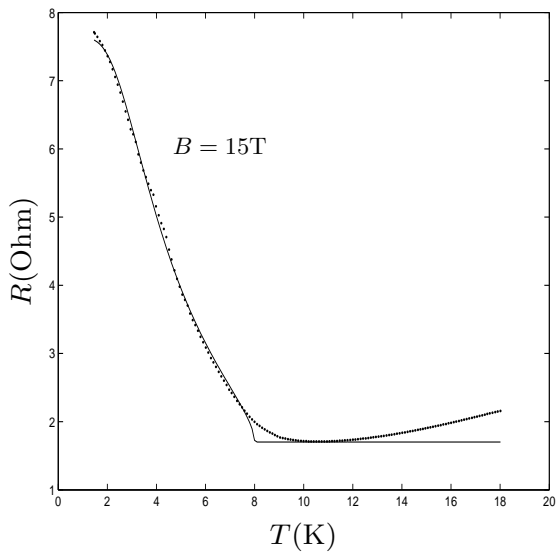}
\caption{The temperature dependent magnetoresistance is shown at $B=15$T. The dots are the experimental data, the
solid line is our fit.}\label{Rpara15T}
\end{figure}

In a magnetic field the quasiparticle spectrum is quantized as first shown by Nersesyan et al.\cite{Ners1,Ners2}.
This dramatic change in the quasiparticle spectrum is most readily seen in ADMR as it has been demonstrated
recently for SDW plus USDW in (TMTSF)$_2$PF$_6$ below $T=T^*(\sim 4$K)\cite{makitmtsf}.
About a decade ago ADMR in LTP in  $\alpha$-(BEDT-TTF)$_2$KHg(SCN)$_4$ salts have been studied intensively. In particular
ADMR for current $\bf j$ perpendicular ($\bf j\parallel b^*$) and parallel to the a-c
plane exhibits a broad
peak around $\theta=0^\circ$ (see insert in
Fig. \ref{fig:koord}),
where $\theta$ is   
the angle with which the magnetic field is tilted from the b-axis (normal to the conducting plane).
In addition, a series of dips are observed at $\theta=\theta_n$ given by\cite{fermi,kovalev}
\begin{equation}
\tan(\theta_n)\cos(\phi-\phi_0)=\tan(\theta_0)+nd_0,
\end{equation}
where $\tan\theta_0\simeq 0.5$, $d_0\simeq 1.25$, $\phi_0\simeq 27^\circ$
and $n=0$, $\pm1$, $\pm2$\dots.
Here $\phi$ is the angle  the projected magnetic field on the a-c plane makes with the c-axis.
The origin of this surprising ADMR have been discussed but apparently without clear answer\cite{fermi,kovalev,
caulfield2,hanasaki,hibas,blundell}.
In the following we shall show that the quasiparticle spectrum in UCDW in $\alpha$-(BEDT-TTF)$_2$KHg(SCN)$_4$ salts
is quantized in the presence of magnetic field.
The small energy gap which is proportional to $\sqrt B$ where B is the field strength, depends also on the direction
of the magnetic field and it can be seen in ADMR. As it will be shown below, we can describe salient aspects of ADMR
seen in LTP of $\alpha$-(BEDT-TTF)$_2$KHg(SCN)$_4$ very consistently.
Therefore we may conclude that ADMR in $\alpha$-(BEDT-TTF)$_2$KHg(SCN)$_4$ provides definitive evidence that LTP
is UCDW.
We stress that the Landau quantization as proposed by Nersesyan et al.\cite{Ners1,Ners2} should be readily accessible in other UCDW
and USDW systems.
In this respect experimental analysis of ADMR in the pseudogap phase in high $T_c$ cuprate superconductors\cite{krakko}
and the glassy phase in $\kappa$-(BEDT-TTF)$_2$Cu[N(CN)$_2$]Br salt\cite{pinteric} will be of great interest.
In $\alpha$-(BEDT-TTF)$_2$KHg(SCN)$_4$ salts the conducting plane is the a-c plane and the quasi-one dimensional Fermi surface 
is perpendicular to the a-axis. In addition there is a quasi-two dimensional Fermi surface with elliptical cross section in the a-c
plane. In LTP we assume that UCDW appears on the quasi-one dimensional Fermi surface with quasiparticle energy given by
\begin{equation}
E({\bf k})=\sqrt{\xi^2+\Delta^2({\bf k})}-\varepsilon_0\cos(2\bf b^\prime k),\label{elso}  
\end{equation}
where $\xi\approx v_a(k_a-k_F)$, $\Delta({\bf k})=\Delta\cos(ck_c)$ and
$\varepsilon_0$ is the parameter describing the imperfect nesting\cite{yamaji1,yamaji2,huang,physicaB}.
In fitting the experimental data we discovered that 1. Eq. (\ref{elso}) gives only one single dip
in ADMR, 2. therefore the imperfect nesting term has to be generalized as
\begin{equation}
\varepsilon_0\cos(2{\bf b^\prime k})\longrightarrow \sum_n\varepsilon_n\cos(2{\bf
b}^\prime_n\bf k),
\label{generalization}
\end{equation}
where ${\bf b}^\prime_n={b}^\prime(\cos\theta_n \hat{\bf
k}_b+\sin\theta_n(\hat{\bf k}_a\cos\phi_0+\hat{\bf k}_c\sin\phi_0))$ and $\varepsilon_n\sim 2^{-|n|}$.
 Eq. (\ref{generalization}) indicates that the imperfect nesting term
does not follow from a usual tight binding model but appears to have
an interesting superstructure whose meaning is not clear at this
moment.              
As seen from Eq. (\ref{elso}), the quasiparticle spectrum is gapless and LTP is metallic in sharp contrast to conventional
CDW.
In a magnetic field the first term of the quasiparticle spectrum changes to
\begin{equation}
E_n=\pm\sqrt{2nv_a\Delta c e
|B\cos\theta|},
\label{ketto}
\end{equation}
where $n=0$, $1$, $2$\dots.
This  is readily obtained following Refs. \onlinecite{Ners1,Ners2}. 
The contribution from the imperfect nesting term is considered as a perturbation and the lowest order corrections to the
energy spectrum are given by:
\begin{gather}
E_0^1=E_1^1=-\sum_m\varepsilon_m\exp(-y_m),\\
E_1^2=-\sum_m\varepsilon_m(1-2y_m)\exp(-y_m),
\end{gather}
where $y_m=v_a {b^\prime}^2e |B\cos(\theta)|(\tan(\theta)\cos(\phi-\phi_o)-(\tan(\theta_0)+md_0))^2/\Delta c$. 
The $n=1$ level was twofold degenerate, but 
the imperfect nesting term splits the degeneracy by $E_1^1$ and $E_1^2$. Also the imperfect nesting term breaks the particle-hole
symmetry.
When $\beta E_1\gg 1$ ($\beta=(k_BT)^{-1}$), the quasiparticle transport in the quasi-one dimensional Fermi surface
is dominated by the quasiparticles at $n=0$ and $n=1$ Landau levels. Considering that there are 2 conducting channels
and only the quasi-one dimensional one is affected by the appearence of UCDW, the ADMR is written as
\begin{eqnarray}
R(B,\theta,\phi)^{-1}=2\sigma_1\left(\dfrac{\exp(-\beta 
E_1)+\cosh(\beta E_1^1)}{\cosh(\beta
E_1)+\cosh(\beta E_1^1)}+\dfrac{\exp(-\beta
E_1)+\cosh(\beta E_1^2)}{\cosh(\beta
E_1)+\cosh(\beta E_1^2)}\right)+\sigma_2
\label{fit}
\end{eqnarray}
Here $\sigma_1$ and $\sigma_2$ are the conductivities of the $n=1$ Landau level and quasi-two dimensional channels, in which the
contribution of the $n=0$ Landau level was melted,
respectively.
In
Figs. \ref{fig:koord} and \ref{Rpara15T} we compare the $B$ dependence of the magnetoresistance at $T=1.4$K and
$T=4.14$K and the $T$ dependence of the magnetoresistance for $B=15T$ for $\theta=0$. In fitting the temperature dependence
of the resistivity, we assumed $\Delta(T)/\Delta(0)=\sqrt{1-(T/T_c)^3}$, which was found to be very close
to the exact solution of $\Delta(T)$\cite{nagycikk}.
             
\begin{figure}[h]
\psfrag{x}[t][b][1][0]{$\theta$ ($^\circ$)}
\psfrag{y}[b][t][1][0]{$R_\parallel(15T,\theta)$ (Ohm)}
\includegraphics[width=7cm,height=7cm]{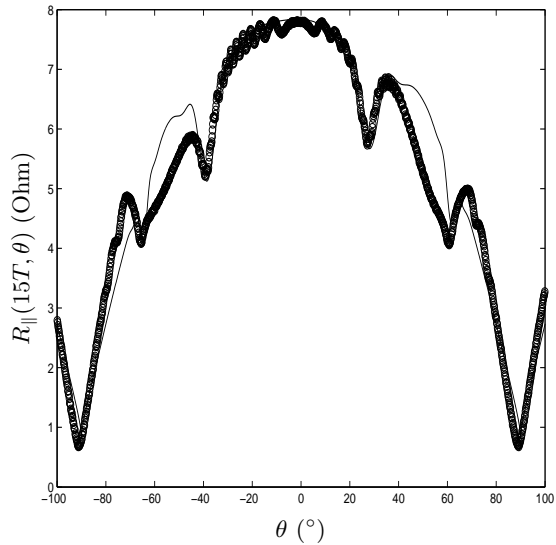}
\caption{The angular dependent magnetoresistance is shown for current parallel
to the a-c plane at $T=1.4$K,
$B=15$T. The open circles belong to the experimental data, the solid
line is our fit based on Eq. (\ref{fit}).}
\label{rpara}
\end{figure}

\begin{figure}[h]
\psfrag{x}[t][b][1][0]{$\theta$ ($^\circ$)}
\psfrag{y}[b][t][1][0]{$R_\perp(15T,\theta)$ (Ohm)}    
\includegraphics[width=7cm,height=7cm]{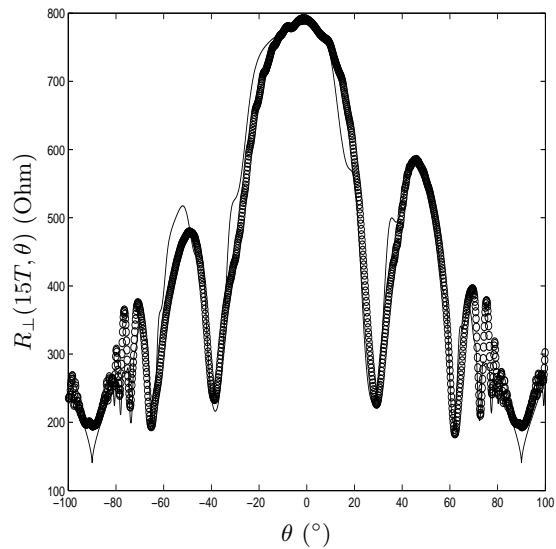}
\caption{The angular dependent magnetoresistance is shown for current perpendicular 
to the a-c plane at $T=1.4$K,
$B=15$T. The open circles belong to the experimental data, the solid
line is our fit from Eq. (\ref{fit}).}
\label{rperp}
\end{figure}

\begin{figure}
\hspace*{-2cm}
\psfrag{x}[t][b][1][0]{$\theta$ ($^\circ$)}
\psfrag{y}[b][t][1][0]{$R_\perp(15T,\theta)$ (Ohm)} 
\includegraphics[width=7cm,height=7cm]{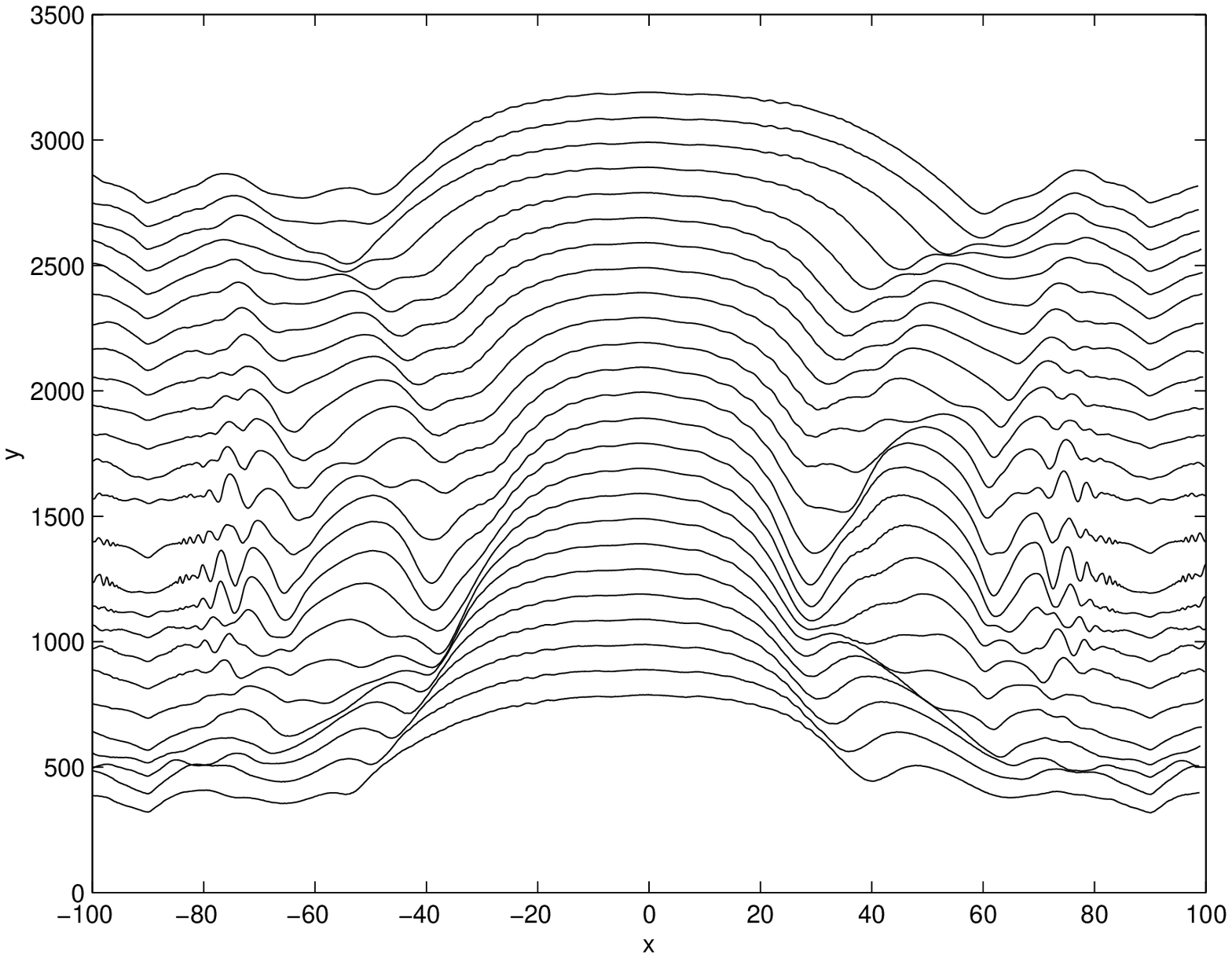}
\hspace*{1cm}
\psfrag{x}[t][b][1][0]{$\theta$ ($^\circ$)}
\psfrag{y}[b][t][1][0]{$R_\perp(15T,\theta)$ (Ohm)} 
\includegraphics[width=7cm,height=7cm]{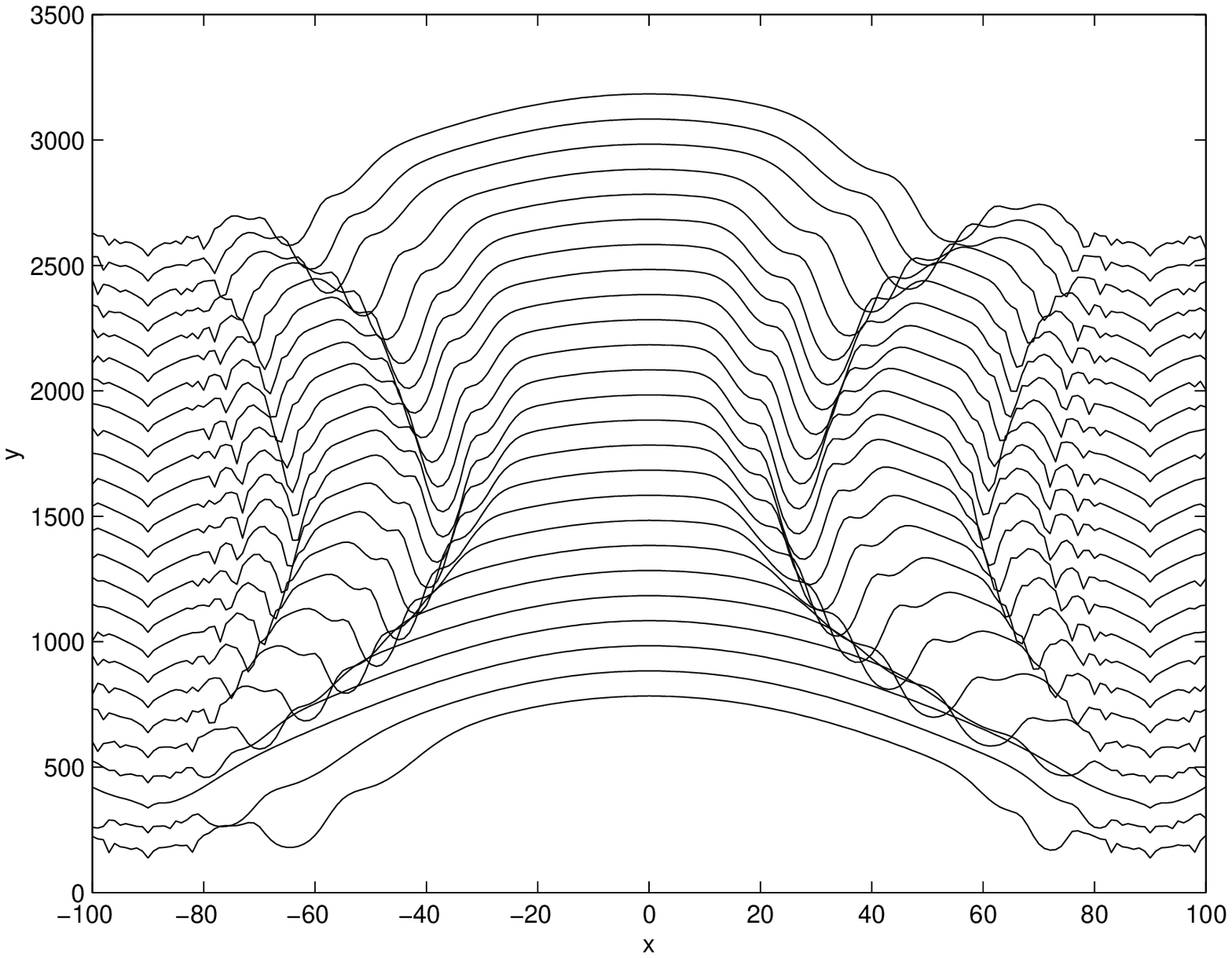}
\caption{ADMR is shown for current perpendicular to the a-c plane at $T=1.4$K and $B=15T$ for $\phi=-77^\circ$, $-70^\circ$,
$-62.5^\circ$, $-55^\circ$, $-47^\circ$, $-39^\circ$, $-30.5^\circ$, $-22^\circ$, $-14^\circ$, $-6^\circ$, $2^\circ$, $10^\circ$,
$23^\circ$, $33^\circ$, $41^\circ$, $48.5^\circ$, $56^\circ$, $61^\circ$, $64^\circ$,
$67^\circ$, $73^\circ$, $80^\circ$, $88.5^\circ$, $92^\circ$ and $96^\circ$ from bottom to top. The left (right) panel shows
experimental
(theoretical) curves, which are shifted from their original position along the vertical axis by $n\times100$Ohm, $n=0$ for $\phi=-77^\circ$, $n=1$ 
for $\phi=-70^\circ$, \dots.}\label{sokphit}
\end{figure}

Clearly the fitting becomes better as $T$ decreases and/or $B$ increases. Also for
$T=1.4$K Shubnikov-de Haas oscillation becomes visible around $B=10$T, then the fitting starts breaking away.
Clearly in this high field region the quantization of Fermi surface itself starts interfering
with the quantization described above. In this region, the explicit $B$ and $T$ dependence of $\sigma_1$
and $\sigma_2$ should be taken into account what we neglected here for simplicity. Also 
the deviaton of the theoretical curve from the experimental one above $T_c$ in Fig. \ref{Rpara15T} is originated
from this neglect. Here we concentrated on the dominant conduction mechanism, that is thermally
excited quasiparticles across the magnetic field induced gap. 
From these fittings we obtain $\sigma_2/\sigma_1$ of the order of $0.1$, and by assuming the
mean field value of $\Delta$ ($17$K), we get $v_a$ of the order of $10^6$cm/s. 
In Figs. \ref{rpara} and \ref{rperp}  we show the experimental data of ADMR as a function of $\theta$ for current parallel and
perpendicular to the conducting plane for $T=1.4$K, $B=15$T and $\phi=45^\circ$.
As is readily seen the fittings are excellent. From this we deduce $\sigma_2/\sigma_1$
of the order of 0.1,
$\varepsilon_0=4.2$K, $b^\prime$ is of the order of a few lattice constants from these fittings.
Finally we show in Fig. \ref{sokphit} $R$ versus $\theta$ for different $\phi$ and compare with the experimental data side by
side. Perhaps there
are still differences in some details but the overall agreement is very striking. These  differences might arise from the fact, that
similarly to the neglect of magnetic field and temperature dependence of $\sigma_1$ and $\sigma_2$, we also assumed them to be
independent of $\phi$ and $\theta$.
The present model can describe a similar figure found in Ref. \onlinecite{hanasaki} rather
well.

In summary we have succeeded in describing the salient feature of ADMR observed in LTP in  
$\alpha$-(BEDT-TTF)$_2$KHg(SCN)$_4$ in term of UCDW with the Landau quantization of the quasiparticle spectrum.
Very similar ADMR have been seen in M=Rb and Tl compounds as well. Therefore we conclude that LTP in
$\alpha$-(BEDT-TTF)$_2$MHg(SCN)$_4$ salts should be UCDW. Also we believe that ADMR provides clear signiture for
the presence of UCDW and USDW. Therefore this technique can be exploited for other possible candidates of UDW.

\begin{acknowledgments}
We are benefitted from discussions with Amir Hamzi\'c, Silvia Tomi\'c and Peter Thalmeier.
One of the authors (K. M.) 
acknowledges the hospitality and support of the Max Planck Institute for the
Physics of Complex Systems, Dresden, where most of this work was done.
This work
was supported by the Hungarian National Research Fund under grant numbers
OTKA T032162 and T037451, and by the Ministry of Education under grant 
number FKFP 0029/1999.
\end{acknowledgments}

\bibliographystyle{apsrev}
\bibliography{mr}
\end{document}